\documentclass[conference]{IEEEtran}

\usepackage{graphicx}
\usepackage{moreverb}
\usepackage{amssymb}
\usepackage{amsfonts}
\usepackage{multicol}
\usepackage{mathtools}
\usepackage[mathscr]{euscript}
\usepackage{booktabs}
\usepackage{makecell}
\usepackage{multirow}
\usepackage{amsbsy} 
\usepackage[hypcap=true, font=small]{caption}
\usepackage[lined,vlined,ruled]{algorithm2e}
\usepackage{cite}
\usepackage{xfrac}
\usepackage{bbm}
\usepackage{eufrak}
\usepackage{color, colortbl}
\definecolor{Gray}{gray}{0.9}
\usepackage[first=0,last=9]{lcg}

\usepackage{authblk}
\usepackage{stfloats}
\usepackage[bottom]{footmisc}

\DeclareMathOperator*{\argmin}{argmin}
\newcommand*{\argminl}{\argmin\limits}
\usepackage{hyperref}
\onecolumn
\begin{document}

\title{Distributed Lifetime Optimization in Wireless Sensor Networks using Alternating Direction Method of Multipliers\vspace{-2ex}}

\author[]{Farzad Tashtarian, Ahmadreza Montazerolghaem, Amir Varasteh}


\maketitle

\thispagestyle{plain}
\pagestyle{plain}
\vspace{2ex}
\begin{abstract}
Due to the limited energy of sensor nodes in wireless sensor networks, extending the networks lifetime is a major challenge that can be formulated as an optimization problem. In this paper, we propose a distributed iterative algorithm based on Alternating Direction Method of Multipliers (ADMM) with the aim of maximizing sensor network lifetime. The features of this algorithm are use of local information, low overhead of message passing, low computational complexity, fast convergence, and consequently reduced energy consumption. In this study, we present the convergence results and the number of iterations required to achieve the stopping criterion. Furthermore, the impact of problem size (number of sensor nodes) on the solution and constraints violation is studied and finally, the proposed algorithm is compared to one of the well-known subgradient-based algorithms.
\end{abstract}


\section{Introduction}
Wireless sensor networks (WSNs) are formed sensor nodes with signal processing power. Recently, these networks have been applied extensively in real-life challenges, such as disaster monitoring, integrated infrastructure monitoring, as well as various military applications \cite{akyildiz2002wireless}. A major limitation of these sensor nodes is that they have limited energy supply and hence limited computing power. Therefore, a serious challenge regarding the WSNs is energy optimization and lifetime maximization which are the main objectives of this study. However, the recent studies on these challenges can be divided into two general categories: centralized and distributed. \par

The adoption of the distributed algorithm is more efficient than the central algorithm in WSNs \cite{madan2006distributed}. That is because, for instance, by increasing the number of the sensor nodes in a WSN, central algorithms cannot be processed on a single node, because it increases the energy consumption and causes network failures and service outages \cite{anastasi2009energy}. At the same time, using distributed algorithms, each of sensor nodes can participate in processing the network information and decision making procedure. Therefore, in this condition, using distributed algorithms would be more energy-efficient and can increase the network lifetime significantly \cite{anastasi2009energy}. \par

There are two general types of distributed algorithms: dual decomposition (subgradient) and ADMM. Most of the distributed methods in this field are based on dual decomposition with subgradient methods \cite{boyd2003subgradient, wei2012distributed}. Dual decomposition with subgradient methods are often used to develop distributed optimization algorithms. They, however, require delicate adjustments of step sizes, which makes convergence difficult to achieve a solution for large-scale problems \cite{he2009distributed}. The best known rate of convergence for subgradient-based distributed methods is much higher than the alternating direction method of multipliers (ADMM) algorithm \cite{wei2012distributed}. Thus, since both approaches are iteration-based and the speed of convergence in ADMM is more than the subgradient method, an ADMM-based distributed algorithm is applicable to increase WSN lifetime. This is because by reducing the number of the iterations to achieve the global solution, the energy consumption in the sensor nodes reduces significantly. In fact, using the distributed ADMM method can reduce the number of messages exchanged between the sensor nodes, which makes this technique even more energy-efficient.\par

The main contribution of this paper is proposing a novel fast distributed algorithm based on the ADMM method with the objective of maximizing WSN lifetime. Such an algorithm can be implemented by parallel execution and the overhead of message passing is low. Our developed algorithm introduces an alternative approach for commonly used dual decomposition. In this paper, first, an optimization problem is formulated, and then, it is solved using the ADMM to achieve fast convergence. There are two main reasons that make the proposed algorithm applicable for energy-efficient implementations. The first reason is only a small amount of local information exchange is required at each iteration. Second, it is capable of generating near-optimal solution in a significantly smaller number of iterations as compared to the state-of-the-art methods based on subgradient. Our algorithm also scales better to large networks and it does not require intensive fine-tuning of the step size. Additionally, our presented approach does not need any central computational node with knowledge about the whole network structure. The algorithm uses only peer-to-peer communications between the neighboring sensor nodes which allows the routing update using only the local information exchange. To establish a benchmark for our distributed algorithm, in performance evaluation section, we compare our work to a subgradient-based algorithm which is provided in \cite{madan2006distributed}. \par

It is worth to mention that from message complexity point of view, both aforementioned distributed algorithms have equal number of transmitted messages at each iteration for a same network topology. However, the message complexity in proposed ADMM-based method is much lower than subgradient method due to the significantly lower number of iterations. In fact, it is proved that the convergence rate of subgradient method is $O(1/\sqrt{k})$ in comparison with $O(1/k)$ in the distributed ADMM \cite{wei2012distributed}. For illustration, we evaluate the number of iterations required to reach near-optimal solution in both algorithms.

The rest of this paper is organized as follows: The related work is described in section II. Next, ADMM and subgradient approaches are presented in section III. The problem formulation and detailed derivation to solve the optimization problem using the distributed ADMM algorithm are proposed in section IV and V, respectively. Also, section VI presents the performance evaluation of the proposed algorithm, and finally, we conclude the paper in section VII.

\section{Related Work}
Over the past years, optimization techniques have been widely used to solve many problems raised in WSNs using either subgradient, or the distributed ADMM methods. In below, we describe the approaches in the literature that have used either of these methods.

\textit{Subgradient-based Algorithms:} Network lifetime maximization for conventional WSNs has been extensively studied using subgradient-based approaches. Authors in \cite{gatzianas2008distributed} presented a distributed algorithm for computing the maximum lifetime of a sensor network which routes data from source nodes to a mobile sink and finally to the main station. In another work \cite{yun2013distributed}, authors proposed a distributed algorithm with the aim of maximizing WSN lifetime. In the mentioned paper, the authors considered a mobile sink to collect the data from source nodes, which the underlying application can tolerate some degree of delay in delivering the data to the sink node. Moreover, in \cite{he2009distributed}, authors derived distributed algorithms using the subgradient method to maximize the wireless visual sensor network lifetime by jointly optimizing the source rates, the encoding powers, and the routing scheme. In \cite{zhu2007tradeoff}, a trade-off between network lifetime maximization and fair rate allocation in sensor networks as a constrained optimization problem with the weighting method was formulated. Then, the original problem was transformed into a convex optimization, which made the subgradient-based optimization approach applicable. Further, authors in \cite{zheng2009joint} studied the trade-off between network utility and network lifetime for energy-constrained WSNs. They provided two algorithms: a partially distributed algorithm and a fully distributed algorithm to tackle the problem.

\textit{ADMM-based Algorithms:} These methods have demonstrated good empirical performance on several distributed applications. Authors in \cite{leinonen2013distributed} proposed a distributed total transmit power minimization in a single-sink data gathering WSN by using the ADMM method. As another application, Liang \textit{et. al.} in \cite{liang2015distributed} developed the ADMM-based distributed wireless virtual resource allocation and in-network caching scheme. Furthermore, developing the application of the ADMM to optimize a dynamic objective function in a decentralized multi-agent system is introduced in \cite{ling2014decentralized}. In addition, a Distributed ADMM named D-ADMM, for solving separable optimization problems in networks of interconnected nodes or agents was proposed in \cite{mota2013d} that focused on a coloring scheme of the network, according to which nodes operate asynchronously.\par 
However, to the best of our knowledge, the distributed ADMM method has not been proposed to maximize WSN lifetime in the existing literature.

\section{Preliminaries: Subgradient and Standard ADMM Algorithms}

\subsection{Subgradient Algorithm}
Subgradient methods are iterative algorithms to solve convex optimization problems \cite{boyd2003subgradient}. Consider the following convex optimization problem:
\begin{align}
    &\textbf{Minimize} \hspace{1cm} w_0(x) \label{eq1}\\
    &\textbf{s.t.} \hspace{1.2cm} w_i(x) \leq 0, \hspace{.3cm} i=1,...,m \nonumber \\
    &\hspace{1.65cm} \mu_i(x) \leq 0, \hspace{.3cm} i=1,...,p \nonumber \\
    & \textbf{vars.} \hspace{.92cm} x \in D \subseteq \mathbb{R}^n \nonumber
\end{align}
where $D$ is a convex set, $w_0(x)$, $w_i$ are convex functions, and $\mu_i$ is a function of $x$. The Lagrangian for $\lambda \succeq 0$, is given by:
\begin{align}
    L(x,\lambda,v) = w_0(x) + \sum_{i=1}^m \nolimits \lambda_i w_i(x) + \sum_{i=1}^p \nolimits v_i \mu_i(x) \label{eq2}
\end{align}
The dual function is $g(\lambda, v)= inf_{x\in D} L(x,\lambda, v)$, and the dual problem is:
\begin{align}
    &\textbf{Maximize} \hspace{1cm} g (\lambda, v)  \nonumber \\
    &\textbf{s.t.} \hspace{1.2cm} \lambda \succeq 0 \nonumber
\end{align}
Also, for $ \lambda \succeq 0$ we have:
\begin{align}
    x^* (\lambda , v) = arg inf_{x \in D} L(x,\lambda , v) \label{eq3}
\end{align}
Initial point is $(\lambda^0,v^0)$ and at each iteration $k$, the following relations are computed:
\begin{align}
    &\lambda_i^{(k+1)} = ( \lambda_i^{(k)} - \alpha_k h_i^{(k)})_+, \text{ } \forall i=1,...,m \label{eq4}\\
    & v_i^{(k+1)} = v_i^{(k)} - \alpha_k f_i^{(k)}, \hspace{.7cm} \forall i=1,...,p \label{eq5}
\end{align}
where $\alpha_k$ is a positive scalar step size and the vector $[ {h^{(k)}}^T , {f^{(k)}}^T ]^T$ is a subgradient of $-g$ at $(\lambda^{(k)} , v^{(k)})$. A subgradient is given by:
\begin{align}
    & h_i^{(k)} = -w_i (x^* (\lambda^{(k)} , v^{(k)} )), \hspace{.2cm} \forall i = 1 , ... ,m \label{eq6}\\
    & f_i^{(k)} = -\mu_i (x^* (\lambda^{(k)} , v^{(k)} )), \hspace{.3cm} \forall i = 1 , ... ,p \label{eq7}
\end{align}
Further, the condition for convergence is:
\begin{align}
    \alpha_k \rightarrow 0, \hspace{.3cm} \sum_{k=1}^{\infty} \nolimits \alpha_k = \infty \label{eq8}
\end{align}
Under the assumptions, the sequence of primal iterates $x^{(k)}=x^*(\lambda^{(k)},v^{(k)} )$ convergence to the near-optimal solution of the primal problem \cite{boyd2003subgradient}.

\subsection{Standard ADMM Algorithm}
The standard ADMM is an augmented Lagrangian based method which solves problems using an iterative primal-dual algorithm as follows:
\begin{align}
    &\textbf{Minimize} \hspace{.7cm} f(y) + g(z)  \\
    &\textbf{s.t.} \hspace{1.2cm} Fy + Dz - c = 0 \nonumber \\
    &\textbf{vars.} \hspace{.95cm} y \in \mathbb{R}^n , z \in \mathbb{R}^m \nonumber
\end{align}
where $f$ and $g$ are convex functions, $c \in \mathbb{R}^p$, and $F$ and $D$ are matrices of dimension $p \times n$ and $p\times m$, respectively (see \cite{boyd2011distributed} for a detailed review). The augmented Lagrangian function of the above model is presented as follows:
\begin{align}
    L_\rho (y,z,\mu) &= f(y) + g(z) + \mu^\intercal (Fy + Dz -c) \nonumber \\
    &+ \frac{\rho}{2} \| Fy + Dz - c \|^2_2 \label{eq9}
\end{align}
where $\mu$ is the Lagrangian multiplier and $\rho$ is the positive penalty scalar. ADMM updates the primal variables $y$ and $z$, followed by a dual variable update $\mu$ in an iterative manner. Considering $(y^0,z^0, \mu^0)$ as the initial vector, the updated variables at iteration $k>0$ are computed as follows:
\begin{align}
    y^k=  \argminl_{y} L_\rho (y,z^{k-1},\mu^{k-1}) \label{eq10}
\end{align}
\begin{align}
    z^k = \argminl_{z} L_\rho (y^k,z,\mu^{k-1}) \label{eq11}
\end{align}
\begin{align}
    \mu^k= \mu^{k-1}-\rho (F y^k +D z^k -c) \label{eq12}
\end{align}
However, to formulate the ADMM into more convenient form by combining the linear and quadratic terms in $L_\rho$, $u = (1/ \rho) \mu$ is defined as the scaled dual variable. Thus, we have:
\begin{align}
    \mu^\intercal (Fy & + Dz - c) + \frac{\rho}{2} \| Fy + Dz - c \|_2^2 \nonumber \\
    &= \frac{\rho}{2} \| Fy + Dz - c + u \|^2_2 - \frac{1}{2\rho} \| u \|^2_2 \label{eq13}
\end{align}
By disregarding independent terms of the minimization variables, the ADMM method can be expressed as:
\begin{align}
    y^k = \argminl_{y} ( f(y) + \frac{\rho}{2} \| Fy+Dz^{k-1} -c + u^{k-1} \|_2^2 ) \label{eq14}
\end{align}
\begin{align}
    z^k = \argminl_{z} ( g(z) + \frac{\rho}{2} \| Fy^k+Dz -c + u^{k-1} \|_2^2 ) \label{eq15}
\end{align}
\begin{align}
    u^k = u^{k-1} + Fy^k + Dz^k -c \label{eq16}
\end{align}
We use the scaled form of ADMM to develop our distributed algorithm. Achieving small upper-bounds of the dual residual $s^{k}$ and the primal residual $r^{k}$ at iteration $k$ can be considered as a reasonable termination criterion for the ADMM algorithm \cite{boyd2011distributed}, i.e.,
\begin{align}
    \|r^{k}\|_2\le\epsilon\text{ and  }\|s^{k}\|_2\le\bar{\epsilon} \label{eq17}
\end{align}
where:
\begin{align}
    &r^{k+1}= Fy^{k+1}+Dz^{k+1} -c \label{eq18} \\
    & s^{k+1}=\rho F^{T}D(z^{k+1}-z^{k}) \label{eq19}
\end{align}

\section{Problem Formulation}

In this study, we modeled a WSN as a directed graph $G=(S,D)$, where $S$ and $D$ are sets of vertices which demonstrate the sensor nodes (including sink node with index 0), and edges representing bi-directional wireless links, respectively. Each sensor node $i$, which is denoted by $S_i$, $i=1:length(S)$, has an initial energy $e_i$ and generates the data traffic at a constant deterministic rate $g_i$ bit per second. Also, let us introduce a cost function $E:D \rightarrow \mathbb{R}^+$ on the set $D$, where $E_{ij}$ indicates the required amount of power to send a composite bit-stream at rate $r_{ij}$ (bps) from $S_i$ to $S_j$ if and only if $d_{ij} \in D$:
\begin{align}
    &C_{ij}= \alpha + \beta d_{ij}^2 \\
    &E_{ij} = C_{ij} r_{ij} \label{eq20}
\end{align}
where the non-negative constant term $\alpha$ depends on the electronics energy which is measured in joule/bit \cite{tashtarian2015maximizing}. We also assume the free space transmission model and use an amplifier energy as $\beta d_{ij}^2$ measured in joule/bit/m$^2$. Also, $d_{ij}$ is the Euclidean distance between $S_i$ and $S_j$. \par

In the following, we initially define the network lifetime $T$ and then we address the problem of maximizing the network lifetime through a distributed optimization approach.

\textbf{Definition 1:} The network lifetime $T$ is defined as the time duration since the network operation is started until the first sensor node drains its entire battery. \par

By defining the lifetime of each sensor node $S_i$ as $\tau_i = \frac{e_i}{\sum_{j \in \mathscr{N}_i} \nolimits C_{ij} r_{ij} }$, the maximum network lifetime equals to $T= min(\tau_i), \forall i \in S$, where $\mathscr{N}_i$ represents the set of $S_i$ neighbour nodes. Therefore, the main problem is how to determine the high-accurate data flow rates, $r_{ij}$, among sensor nodes with respect to the fixed location of the sink and satisfying the following constraints $\forall i \in S$:
\begin{align}
    &\sum_{j \in \mathscr{N}_i} \nolimits (r_{ij} - r_{ji}) = g_i \label{eq21}\\
    T &\sum_{j \in \mathscr{N}_i} \nolimits C_{ij} r_{ij} \leq e_i  \label{eq22}
\end{align}
The first constraint states that the sum of total incoming flow rates plus self-generated data rate $g_i$ must be equal to the sum of total outgoing flow rates to other nodes, $r_{ij}$, including sink with index $0$ ($g_0=-\sum_{(i=1:length(S))} \nolimits g_i $). The second constraint (Eq. (\ref{eq22})) ensures that the total required transmission energy 
does not exceed its initial energy $e_i$. Therefore, the mathematical optimization model to obtain the optimal data rates can be presented as follows:
\begin{align}
    &\textbf{Maximize} \hspace{1cm} T \label{eq23} \\
    &\textbf{s.t.} \hspace{1.1cm} \text{ constraints (\ref{eq21}) and (\ref{eq22})} \nonumber \\
    & \text{\textbf{vars.} \hspace{.7cm} $T$, $r_{ij} \geq 0$} \nonumber
\end{align}
We note that the definition of network lifetime, and the mathematical model (\ref{eq23}) have been used in some of the previous studies \cite{madan2006distributed}. However, our main contribution is to solve the model (\ref{eq23}) using the distributed ADMM method.

\section{Proposed Distributed ADMM-Based Algorithm}
Our main motivation of proposing a distributed algorithm based on ADMM is that each sensor node $i$ computes its transmission data rate while all nodes converge to an identical maximum network lifetime in a reasonable time. Further, message complexity or the number of transmitted messages among nodes to converge to the global value is one of the main issues that must be taken into account. To design a distributed ADMM-based algorithm for mathematical model (\ref{eq23}), firstly, we convert the nonlinear model (\ref{eq23}) into a linear programming model by introducing $q=1/T$. Thus, we have:
\begin{align}
    &\textbf{Minimize} \hspace{1cm} q \label{eq24} \\
    &\textbf{s.t.} \hspace{1.1cm} \text{ constraints (\ref{eq21}) and (\ref{eq22})} \nonumber \\
    & \textbf{vars.} \hspace{.95cm} q \text{ and } r_{ij} > 0 \nonumber
\end{align}

\begin{figure}[b]
\centering
\includegraphics[width=.63\linewidth]{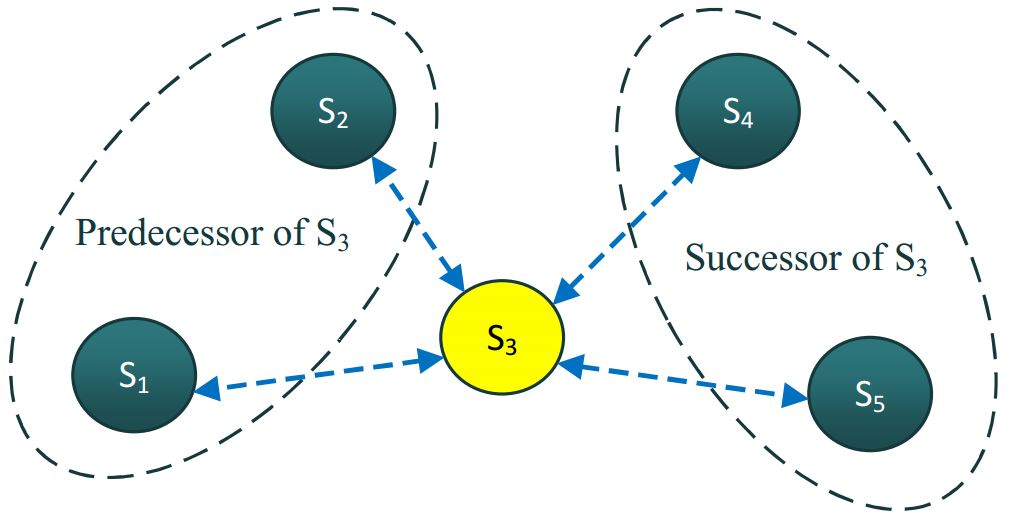}
\caption{The successor and predecessor sets of $S_3$}
\label{fig0}
\end{figure}

To simplify the development of our distributed ADMM-based model, we inspire the proposed algorithm in \cite{wei2012distributed}, and divide the neighbors of $S_i$ into two sets: predecessors and successors of $S_i$, denoted by $\mathbb{P}(i)$ and $\mathbb{S}(i)$, respectively. If $i<j$ and $d_{ij}\in D$, then $S_i \in \mathbb{P}_j$ and $S_j \in \mathbb{S}_i$. Similarly, if $i>j$ and $d_{ij} \in D$, then $S_j \in \mathbb{P}_i$ and $S_i \in \mathbb{S}_j$ (see Fig. \ref{fig0}). In addition, casting the inequality constraint (\ref{eq22}) to an equality form requires us to introduce some auxiliary variables to facilitate the using of ADMM method.\par
Let $q_i$ be the local network lifetime obtained in $S_i$. To achieve the global network lifetime, we should set $q_i=q_j \text{, } \forall i,j \in S$. Moreover, we define a positive variable $z_i$ to convert inequality constraint (\ref{eq22}) to the following equality form:
\begin{align}
    \sum_{j \in \mathscr{N}_i} \nolimits C_{ij} r_{ij} = (q_i - z_i) e_i \text{,   } \forall i \in S \label{eq25}
\end{align}
Since each node $i$ should determine $r_{ij}$ in a distributed manner, $A_{ij}$ is defined as the difference between $r_{ij}$ and $r_{ji}$, $\forall j \in \mathscr{N}_i$; and substitute the first constraint (i.e. Eq. (\ref{eq21})) in model (\ref{eq23}) with:
\begin{align}
    & r_{ij} - r_{ji} = A_{ij} \text{, } \hspace{.6cm}\forall i \in S, j \in \mathscr{N}_i \label{eq26}\\
    & \sum_{j \in \mathscr{N}_i} \nolimits A_{ij} = g_i\text{, } \hspace{.5cm}\forall i \in S \label{eq27}
\end{align}
Putting all together, we define the following convex model with equality constraints to propose a distributed ADMM-based algorithm:
\begin{align}
    &\label{eq28} \textbf{Minimize} \hspace{2cm} \sum_{i \in S} \nolimits q_i  \\
    &\textbf{s.t.} \hspace{1cm} r_{ij} - r_{ji} = A_{ij} \text{, } \hspace{1.9cm}\forall i \in S, j \in \mathscr{N}_i \hspace{.2cm} \textbf{(I)}\nonumber \\
    &\hspace{1.4cm} \sum_{j \in \mathscr{N}_i} \nolimits A_{ij} = g_i \text{, } \hspace{2cm}\forall i \in S \hspace{.85cm} \textbf{(II)} \nonumber \\
    &\hspace{1.4cm} \sum_{j \in \mathscr{N}_i} \nolimits C_{ij} r_{ij} = (q_i - z_i) e_i, \hspace{.45cm}\forall i \in S \hspace{.75cm} \textbf{(III)}\nonumber \\
    &\hspace{1.5cm} q_i - q_j =0 \hspace{3cm} \forall i,j \in S \hspace{.5cm} \textbf{(IV)} \nonumber \\
    & \textbf{vars.} \hspace{.8cm} q_i, z_i,r_{ij},\text{ and } A_{ij}\geq 0 \nonumber
\end{align}
Now, we are able to solve the proposed model (\ref{eq28}) in a distributed way. The augmented Lagrangian function of model (\ref{eq28}) is as follows:
\begin{align}
    L_\rho ( r&,q,z,A,\lambda, \mu, \gamma , \varphi) = \sum_{i \in S} \nolimits q_i \nonumber \\
    & + \lambda_{ij} (r_{ij} - r_{ji} - A_{ij}) + \frac{\rho}{2} \| r_{ij} - r_{ji} - A_{ij}\|_2^2  \nonumber \\
    & + \mu_i (\sum_{j \in \mathscr{N}_i} \nolimits A_{ij} - g_i ) + \frac{\rho}{2} \| \sum_{j \in \mathscr{N}_i} \nolimits A_{ij} - g_i \|_2^2 \nonumber \\
    & + \gamma_i (\sum_{j \in \mathscr{N}_i} \nolimits C_{ij} r_{ij} - (q_i - z_i) e_i) \nonumber \\
    &+ \frac{\rho}{2} \| \sum_{j \in \mathscr{N}_i} \nolimits C_{ij} r_{ij} - (q_i - z_i) e_i \|_2^2 \nonumber \\
    &+ \varphi_{ij} (q_i - q_j) + \frac{\rho}{2} \| q_i - q_j \|_2^2 \label{eq29}
\end{align}

\noindent Let us define $x_{ji}^{(k\pm)}$ where if $j \in \mathbb{P}$, then $x_{ji}^{(k\pm)}$= $x_{ji}^{(k)}$, and if $j \in \mathbb{S}$, then $x_{ji}^{(k\pm)}$= $x_{ji}^{(k-1)}$. Now, each node $i$ must compute $r_{ij}^k$, $q_i^k$, $z_i^k$ and $A_{ij}^k$ at iteration $k$ in a sequential order as follows:
\begin{align}
    r_{ij}^{(k)} &= argmin \big( \frac{\rho}{2} \| r_{ij} - r_{ji}^{(k\pm)} - A_{ij}^{(k-1)} + \frac{1}{\rho} \lambda_{ij}^{(k-1)} \|^2 \nonumber \\ 
    & + \frac{\rho}{2} \| r_{ji}^{(k\pm)} - r_{ij}- A_{ji}^{(k\pm)} + \frac{1}{\rho} \lambda_{ji}^{(k-1)} \|^2 \nonumber \\
    & + \frac{\rho}{2} \| C_{ij} r_{ij}+ \sum_{l \in \mathscr{N}_i, l \neq j} \nolimits  C_{il} {r}_{il}^{(k\pm)} - (q_i^{(k-1)} - z_i^{(k-1)} ) e_i \nonumber\\
    &+ \frac{1}{\rho} \gamma_i^{(k-1)} \|^2  \big) \label{eq30}
\end{align}
\begin{align}
    q_i^{(k)} &= argmin \big( q_i \nonumber \\
    &+ \frac{\rho}{2} \| \sum_{j\in \mathscr{N}_i} \nolimits C_{ij} r_{ij}^{(k)} - (q_i - z_i^{(k-1)}) e_i  + \frac{1}{\rho} \gamma_i^{(k-1)} \|^2 \nonumber \\ &+ \frac{\rho}{2} \sum_{j\in \mathscr{N}_i} \nolimits \| q_i - q_j^{(k\pm)} + \frac{1}{\rho} \varphi_{ij}^{(k-1)} \|^2  \big) \label{eq31}
\end{align}
\begin{align}
    z_i^{(k)} =& argmin \big(\frac{\rho}{2} \| \sum_{j \in \mathscr{N}_i} \nolimits C_{ij} r_{ij}^{(k)} - (q_i^{(k)} - z_i)e_i \nonumber\\
    &+ \frac{1}{\rho} \gamma_i^{(k-1)} \|^2 \big) \label{eq32}
\end{align}
\begin{align}
     A_{ij}^{(k)} &= argmin \big( \frac{\rho}{2} \| r_{ij}^{(k)} - r_{ji}^{(k\pm)} - A_{ij} + \frac{1}{\rho} \lambda_i^{(k-1)} \|^2  \nonumber \\
     &+ \frac{\rho}{2} \| A_{ij} + \sum_{l \in \mathscr{N}_i, l \neq j} \nolimits A_{il} - g_i+ \frac{1}{\rho} \mu_i^{(k-1)} \|^2     \big)  \label{eq33}
\end{align}
Moreover, the dual variables: $\lambda_{ji}^k$, $\varphi_{ji}^k$, $\gamma_i^k$, and  $\mu_i^k$ for each node $i$ must be updated at iteration $k$:
\begin{align}
    \lambda_{ji}^{(k)} = \lambda_{ji}^{(k-1)} + \rho ( r_{ji}^{(k)} - r_{ij}^{(k)} - A_{ji}^{(k)} ), \forall j \in \{\mathscr{N}_i, \mathbb{P}(i)\} \label{eq34}
\end{align}
\begin{align}
   \varphi_{ji}^{(k)} = \varphi_{ji}^{(k-1)} + \rho ( q_{j}^{(k)} - q_{i}^{(k)}), \forall j \in \{\mathscr{N}_i, \mathbb{P}(i)\} \label{eq35}
\end{align}
\begin{align}
    \gamma_i^{(k)} = \gamma_i^{(k-1)} + \rho \big( \sum_{j \in \mathscr{N}_i} \nolimits E_{ij} r_{ij}^{(k)} - (q_i^{(k)} - z_i^{(k)})e_i \big) \label{eq36}
\end{align}
\begin{align}
    \mu_i^{(k)} = \mu_i^{(k-1)} + \rho \big( \sum_{j \in \mathscr{N}_i} \nolimits A_{ij}^{(k)} - g_i \big) \label{eq37}
\end{align}

\begin{figure}[t]
\centering
\includegraphics[width=0.64\linewidth]{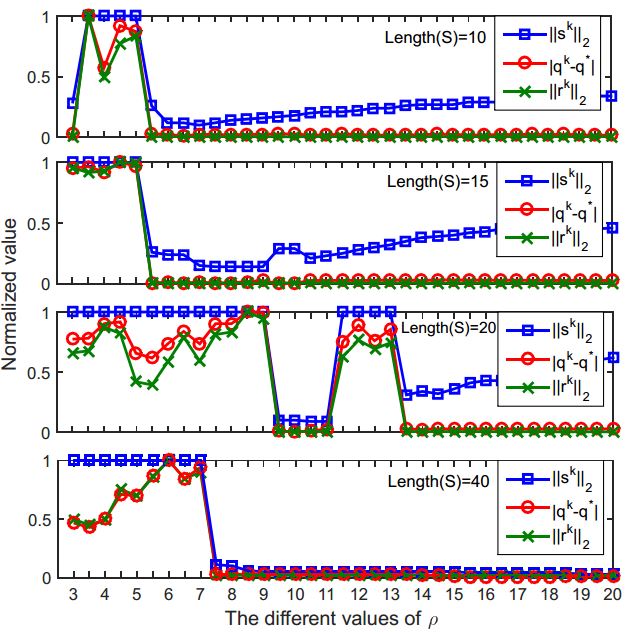}
\caption{Results for $\rho$ estimation in different network sizes}
\label{fig1}
\end{figure}
\begin{figure}[b]
\centering    
\includegraphics[width=.63\linewidth]{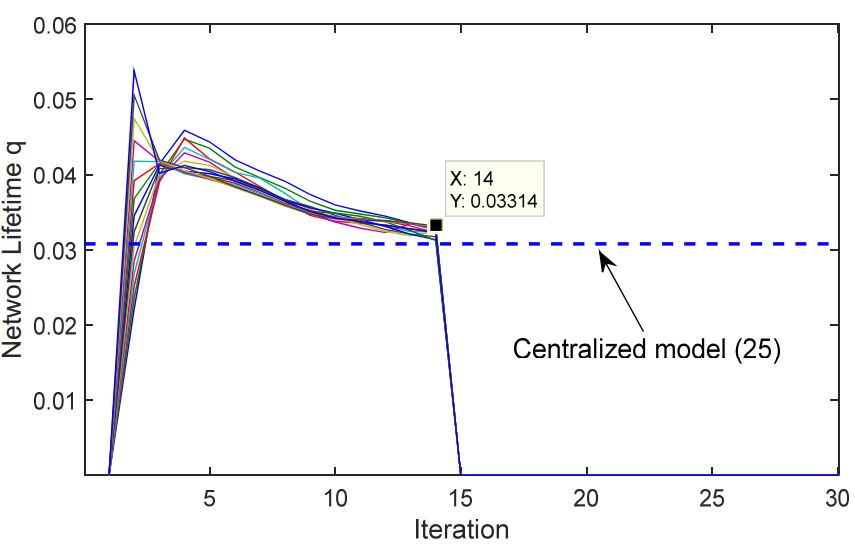}
\caption{ADMM convergence to global solution at $14^{th}$ iteration}
\label{fig2}
\end{figure}
\begin{figure}[t]
  \centering    
  \includegraphics[width=.63\linewidth]{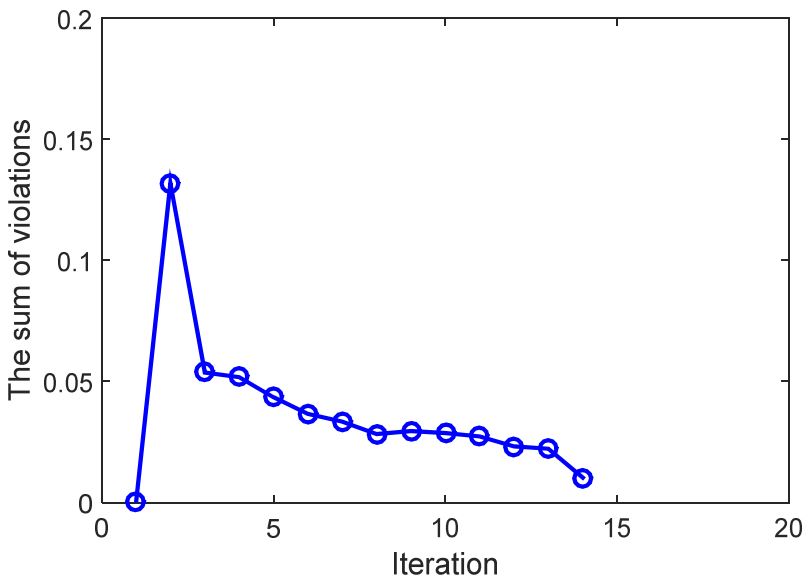}
  \caption{Total violation}
  \label{fig3}
\end{figure}
\begin{figure}[t]
  \centering    
  \includegraphics[width=.62\linewidth]{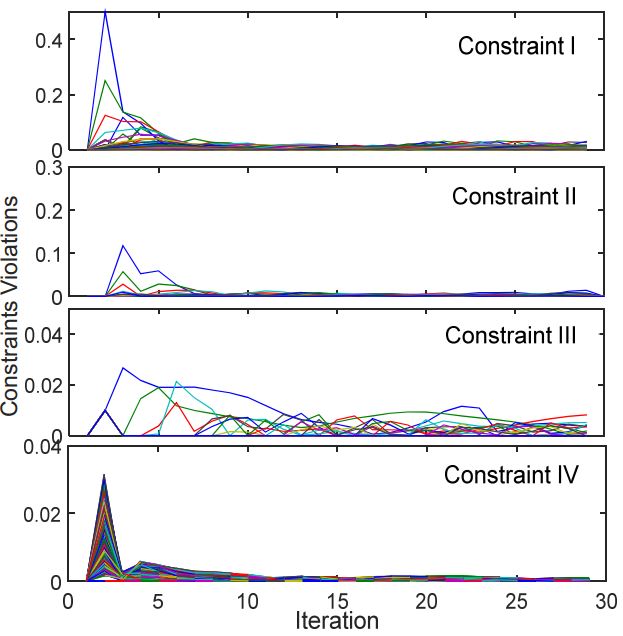}
  \caption{The violation of constrains in model (\ref{eq28})}
  \label{fig4}
\end{figure}
The pesudo code of the proposed algorithm is illustrated in Algorithm \ref{alg1}. In the proposed ADMM algorithm, it is assumed that as soon as node $i$ updates its primal variables $r^{(k)}_{ij}$, $A^{(k)}_{ij}$, and also the dual variables $\lambda^{(k)}_{ji}$ and $\varphi^{(k)}_{ji}$, $S_i$ sends them to all of its neighbours (see Fig. \ref{fig0}). For example, in $k^{th}$ iteration, $S_i$ can update $r^{(k)}_{ij}$ (Eq. (\ref{eq30})) by knowing the primal and dual variables of its neighbors.
\begin{algorithm}[t]
 \caption{The proposed  distributed ADMM algorithm}
 \label{alg1}
\SetAlgoLined
 Initialization: select some arbitrary values for $r_{ij}^0$, $q_i^0$, $z_i^0$, $A_{ij}^0$, $\lambda_{ij}^0$, $\mu_i^0$, $\gamma_i^0$, and $\varphi_{ij}^0$, $\forall i \in S, \text{ and } j \in \mathscr{N}_i$\;
 \For{ each iteration $k$}{
    \For{ each node $i$ } {
    update $r_{ij}^k$, $q_i^k$, $z_i^k$, $A_{ij}^k$, $\forall j \in \mathscr{N}_i$ using Eq. (\ref{eq30})-(\ref{eq33}), respectively\;
    update $\lambda_{ji}^k$, $\varphi_{ji}^k$,  $\gamma_i^{k}$, $\mu_i^k$,  $\forall j \in \{ \mathscr{N}_i\text{, }\mathbb{P}(i)\}$ using Eq. (\ref{eq34})-(\ref{eq37}), respectively\;
    }
  }
\end{algorithm}
\begin{figure}[t]
  \centering    
  \includegraphics[width=.63\linewidth]{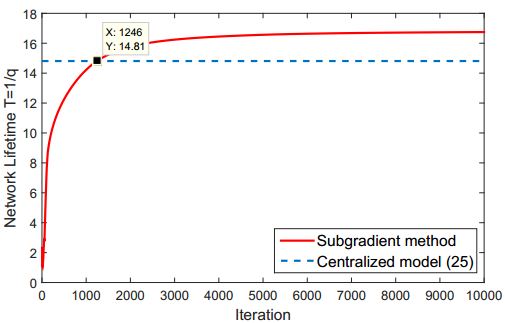}
  \caption{Convergence to global solution in subgradient method at iteration 1246}
  \label{fig5}
\end{figure}

\textbf{Theorem 1:} Considering $\sum_{i \in  S} \nolimits q_i^2$ as the objective function in model (\ref{eq28}) and also modifying Eq. (\ref{eq31}) as:
\begin{align}
    q_i^{(k)} &= argmin \big( q_i^2 \nonumber \\
    &+ \frac{\rho}{2} \| \sum_{j\in \mathscr{N}_i} \nolimits C_{ij} r_{ij}^{(k)} - (q_i - z_i^{(k-1)}) e_i  + \frac{1}{\rho} \gamma_i^{(k-1)} \|^2 \nonumber \\ &+ \frac{\rho}{2} \sum_{j\in \mathscr{N}_i} \nolimits \| q_i - q_j^{(k\pm)} + \frac{1}{\rho} \varphi_{ij}^{(k-1)} \|^2  \big)\label{eq38}
\end{align}
the proposed ADMM algorithm can converge to the high-accurate solution.
\begin{proof}
Since the considered objective function is strongly convex, and assuming a constant penalty value $\rho$ for all iterations, the proposed model (\ref{eq28}) converges to the high-accurate solution \cite{boyd2011distributed,tsianos2011distributed}. 
\end{proof}

\section{Performance Evaluation}
In this section, numerical results are provided to show the performance of the proposed ADMM algorithm. We also compare it to the subgradient-based approach presented in \cite{madan2006distributed}. We use random topologies with different number of sensor nodes length$(S) =10, 15, 20, \text{ and } 40$ in a circular area with the radius equals to 100 units. Our experiments are simulated in MATLAB software and run on a computer equipped with an Intel core i7 2.5 GHz processor and 8 GB of memory. In the following, we firstly demonstrate the impact of different $\rho$ values on stopping criteria and the performance of ADMM algorithm. Secondly, the obtained results from different number of sensor nodes are presented, and then, the constraints violation values for length$(S) = 15$ are analyzed. Finally, the results of the subgradient method presented in \cite{madan2006distributed} are compared to the proposed ADMM-based method.\par

As mentioned earlier, the proposed algorithm uses a constant penalty value $\rho$. To estimate an appropriate $\rho$ with respect to the network parameters, we run the algorithm (in 200 iterations) for different $\rho$ values and measure the quality of solutions, and also $\|r^k\|_2$ and $\|s^k\|_2$ (see Eq. (\ref{eq18}), (\ref{eq19})). 
The normalized obtained values are shown in Fig. \ref{fig1} for different length of $S$. In case of length$(S) = 10$, the best solution is achieved by considering $\rho = 7$. Furthermore, the difference of the distributed ADMM-based algorithm solution and the optimal solution is almost zero $(|q^k-q^* |)$. Besides, when length$(S) = 10$ and $\rho = 7$, the lowest value is obtained for the stopping criteria and $|q^k-q^* |$ also has the lowest values. Therefore, the best solution for length$(S) = 10$ is obtained by using $\rho = 7$. A remarkable note is that the stopping criterion and solution are attainable for every network sizes.

In Fig. \ref{fig2}, the horizontal axis represents the number of iterations and the vertical axis shows the network lifetime $q$ for a sample topology with 15 sensor nodes. In this figure, the blue dotted line represents the solution obtained in the central state and the rest of the curves illustrate the
obtained network lifetime $q_i$ by $S_i$ in a distributed ADMM-based manner. As it is depicted in this figure, the achieved local network lifetime $q_i$ through running ADMM algorithm could converge to the near-optimal solution after $14$ iterations. In fact, the ADMM algorithm stops at $14^{th}$ iteration because of the satisfied stopping conditions defined in Eq. (\ref{eq18}) and (\ref{eq19}) (we set $\epsilon=0.01 \text{ and }\bar{\epsilon}=0.01$). The low number of the iterations has an undeniable influence on reducing the energy consumption and increasing the sensor nodes lifetime. \par

As our next experiment, Fig. \ref{fig3} shows the total violation of the proposed model (\ref{eq28}) for length$(S)=15$. As it can be seen in this figure, the violation reaches to a minimum amount by increasing the iterations, e.g., the total violation at iteration 14 reaches to 0.01. Fig. \ref{fig4} shows the violation of constraints (I) to (IV) in model (\ref{eq28}) which are reduced to less than 0.005 at $14^{th}$ iteration. These results indicate the high convergence speed of the proposed method. To compare the results of the proposed ADMM-based method, in this part, the proposed model (\ref{eq24}) is stimulated using the subgradient method presented in \cite{madan2006distributed} for an identical topology with 15 sensor nodes. 

In Fig. \ref{fig5}, the horizontal axis represents the number of iterations and the vertical axis shows the network lifetime. As it is shown in this figure, the number of required iterations to achieve the global solution in the subgradient method is 1,246 which is significantly higher than the ADMM-based method (which is 14 as it is shown in Fig. \ref{fig2}). Consequently, in subgradient method, the energy consumption of sensor nodes increases and the lifetime of the sensor nodes reduces accordingly. Due to the space limitation, the impact of message complexity on the energy consumption of the sensor nodes is omitted.


\section{Conclusion}
In the past few years, energy efficiency in the Wireless Sensor Networks (WSNs) has become a major challenge in both academia and industry. That is because, each sensor node has a limited energy supply. Therefore, network energy optimization and lifetime maximization are serious problems in WSNs to explore. Due to the nature of this problem, there has been wide attention to distributed optimization approaches. One of these distributed approaches is Alternating Direction Method of Multipliers (ADMM) which has demonstrated a great empirical performance on several distributed applications. However, this method has not been used so far in order to maximize the WSN lifetime. Due to the limited energy consumption in sensor nodes and low computational complexity of the ADMM, the applications of this method in WSNs have had impressive results. Therefore, in this paper, we presented a distributed iterative algorithm based on the ADMM to increase the WSN lifetime by determining the high-accurate transmission rates from sensor nodes to the fixed sink node. In fact, in this approach, the problem decomposed into locally solvable per-node sub-problems which require a small amount of local information exchange. Extensive numerical results illustrated how significantly faster the proposed algorithm can converge to the near-optimal solution comparing to commonly used subgradient-based methods. \par
As future work directions, there are several challenges which are worth to explore. One of these challenges is applying the distributed ADMM optimization in other WSN applications, like WSNs with a mobile sink to collect data from source nodes. In addition, different parameters such as packet loss, node/link failures, delay, etc. can be considered as constraints to enhance the optimization problem. As the last direction, proposing an asynchronous ADMM-based algorithm for WSN lifetime maximization is an interesting problem to discover.

\bibliographystyle{unsrt}
\bibliography{main.bib}

\end{document}